\documentclass{webofc}
\usepackage[varg]{txfonts}   
\usepackage{hyperref}
\hypersetup{
    colorlinks=true,
    linkcolor=black,
    filecolor=magenta,      
    urlcolor=blue,
    citecolor=blue,
}

\begin{document}
\title{The U.S. CMS HL-LHC R\&D Strategic Plan}
%
%

\author{
    \firstname{Oliver} \lastname{Gutsche}\inst{1}\fnsep \and
    \firstname{Tulika} \lastname{Bose}\inst{2}\fnsep \and
    \firstname{Margaret} \lastname{Votava}\inst{1}\fnsep \and
    \firstname{David} \lastname{Mason}\inst{1}\fnsep \and
    \firstname{Andrew} \lastname{Melo}\inst{3}\fnsep \and
    \firstname{Mia} \lastname{Liu}\inst{4}\fnsep \and
    \firstname{Dirk} \lastname{Hufnagel}\inst{1}\fnsep \and
    \firstname{Lindsey} \lastname{Gray}\inst{1}\fnsep \and
    \firstname{Mike} \lastname{Hildreth}\inst{5}\fnsep \and
    \firstname{Burt} \lastname{Holzman}\inst{1}\fnsep \and
    \firstname{Kevin} \lastname{Lannon}\inst{5}\fnsep \and
    \firstname{Saba} \lastname{Sehrish}\inst{1}\fnsep \and
    \firstname{David} \lastname{Sperka}\inst{6}\fnsep \and
    \firstname{James} \lastname{Letts}\inst{7}\fnsep \and
    \firstname{Lothar} \lastname{Bauerdick}\inst{1}\fnsep \and
    \firstname{Kenneth} \lastname{Bloom}\inst{8}\fnsep
}

\institute{
    Fermi National Accelerator Laboratory \and
    University of Wisconsin-Madison \and
    Vanderbilt University \and
    Purdue University \and
    Notre Dame University \and
    Boston University \and
    UC San Diego \and
    University of Nebraska-Lincoln
}

\abstract{The HL-LHC run is anticipated to start at the end of this decade and will pose a significant challenge for the scale of the HEP software and computing infrastructure. The mission of the U.S. CMS Software \& Computing Operations Program is to develop and operate the software and computing resources necessary to process CMS data expeditiously and to enable U.S. physicists to fully participate in the physics of CMS. We have developed a strategic plan to prioritize R\&D efforts to reach this goal for the HL-LHC. This plan includes four grand challenges: modernizing physics software and improving algorithms, building infrastructure for exabyte-scale datasets, transforming the scientific data analysis process and transitioning from R\&D to operations. We are involved in a variety of R\&D projects that fall within these grand challenges. In this talk, we will introduce our four grand challenges and outline the R\&D program of the U.S. CMS Software \& Computing Operations Program.}
\maketitle
\section{Introduction}
\label{sec:intro}

The Compact Muon Solenoid (CMS)~\cite{cms} experiment at the Large Hadron Collider (LHC)~\cite{1748-0221-3-08-S08001} has had a very successful physics program so far with over 1200 scientific papers submitted to date~\cite{cmspubbytime}. The success of this physics program has been enabled by the availability of sufficient computing resources to store, process and analyze the data in an efficient fashion. The CMS experiment is designed, built, and operated by a collaboration of close to 200 institutions across more than 50 countries, and comprises roughly 3,000 members, of which close to 2/3 are physicists with authorship privileges on all CMS physics papers~\cite{cmspeople}. 

The U.S. makes up about 30\% of the authors across a total of $\sim$50 institutions. Both the U.S. Department of Energy (DOE)~\cite{doe} and the U.S. National Science Foundation (NSF)~\cite{nsf} are supporting research at these universities. The U.S. funding agencies centrally support  both the U.S. contributions to the construction of the CMS detector components, and the operation and maintenance of these detector components and U.S. contributions to the software and computing needs of CMS. The latter is organized within the U.S.~CMS Operations Program. The U.S.~CMS Software \& Computing (S\&C) Operations Program~\cite{uscmssnc} is managing the software and computing parts and is referred to as the U.S.~CMS Operations Program in the following.

The U.S.~CMS Operations Program is responsible for supporting U.S. physicists to effectively participate in the LHC scientific program. Amongst its responsibilities are:
\begin{enumerate}
\item
  Provide, maintain, upgrade, and operate the integrated computational and data infrastructure and software services to support CMS physics data analysis in the U.S., across a distributed high-throughput computing system including the Fermilab facilities and seven university centers.
\item
  Support key areas of software development, to maintain and upgrade core software systems, workflow and data management services, and support the modernization of physics software.
\end{enumerate}

Because of the importance of software and computing to the overall physics output of the experiment, the U.S.~CMS Operations Program has historically made significant investments in the development and operation of software tools and computing facilities. DOE and NSF-supported efforts are central to CMS computing. DOE-supported Fermilab~\cite{fnal} provides CMS's major computing facility outside of CERN. Fermilab has core expertise and provides development efforts for software frameworks and libraries, data management software, resource provisioning software and cybersecurity. NSF support is critical for operating sites at 7 Universities in the U.S. These facilities form the backbone of the sites supporting CMS and host official CMS datasets, both real and simulated, and a significant amount of user data. NSF-funded institutes play major roles in algorithmic software development, workflow management software, data management software and specialized analysis environments. DOE and NSF funds are also used for supporting operations of the U.S.~CMS and CMS software and computing infrastructure and HL-LHC R\&D activities. The High Luminosity LHC (HL-LHC) promises scientific results of even great impact, currently out of reach of the LHC, thanks to the upgrade in instrumentation, and a much greater beam intensity.

\subsection{The Scale of HL-LHC Computing and Estimated CMS Resource Needs}
\label{subsec:scale}

During the 15 year lifetime of the LHC program, software and computing needs have posed a number of significant and increasing challenges, caused by the distributed nature of the integrated computational and data infrastructure, and the steadily increasing data complexity and scale due to continuous advances in LHC instantaneous luminosity and thus increasing particle multiplicities, occupancies, and additional proton-proton interactions per collision. While LHC computing demands and the scale of the computing problem have continuously increased in the past, CMS and the U.S.~CMS Operations Program have responded successfully with an ongoing S\&C development program to address these challenges.


HL-LHC pushes these boundaries again, in computing scale, compared to today. HL-LHC, which will take data in several running periods from 2029 to 2042~\cite{bib:LHCschedule}, will collect over 10 times the data collected so far at the LHC. Higher intensity proton-proton collisions and new CMS detector components with higher granularity and more channels will increase the challenges for software and computing significantly:
\begin{enumerate}[(1)]
\item
  The number of events to be processed each year is expected to be larger by a factor of three compared to LHC: 150 Billion events per year.
\item
  Data sample sizes will be larger by a factor of five: the projected total disk storage needs of CMS by 2030 approaches 1/2 exabyte.
\item
  Most data are active and need to be held on quasi-randomly accessible storage systems to be processed by hundreds of simultaneous processing pipelines. This requires very large active storage systems for "hot" data, in addition to agile access to "cold" archival storage.
\item
  Large and highly granular databases with temporal dependencies are used for geometry, calibration, run conditions, and other meta-data.
\item
  Physics software in 2023 consists of more than 4 million lines of highly specialized code that features high algorithmic complexity (e.g.\ manipulation of small matrices for tracking, complex geometry for clustering, etc.) and low computational intensity, thus providing unique challenges to efficiently match code and data throughput to industry-standard accelerators.
\end{enumerate}

CMS has documented~\cite{bib:cmsofflinecomputingresults} its current best estimates of computing needs for the HL-LHC era and the computing model and assumptions made for these resource need predictions. As part of the reviewing process for the HL-LHC detector upgrades, CMS has developed a set of scenarios~\cite{Software:2815292} that should provide the necessary computing resources for the CMS HL-LHC program at a cost similar to that of the computing resource needs for the current program. It also demonstrates that a "build-to-cost" approach can be taken to mitigate risks: in case of less available resources the experiment can partially compensate by increasing physics thresholds e.g. for reconstruction algorithms, and thus initially compromising physics sensitivities, while eventually being able to catch up when additional computing resources should become available e.g. for re-processing data.

\begin{figure}[!tb]
\centering
\includegraphics[width=0.48\textwidth]{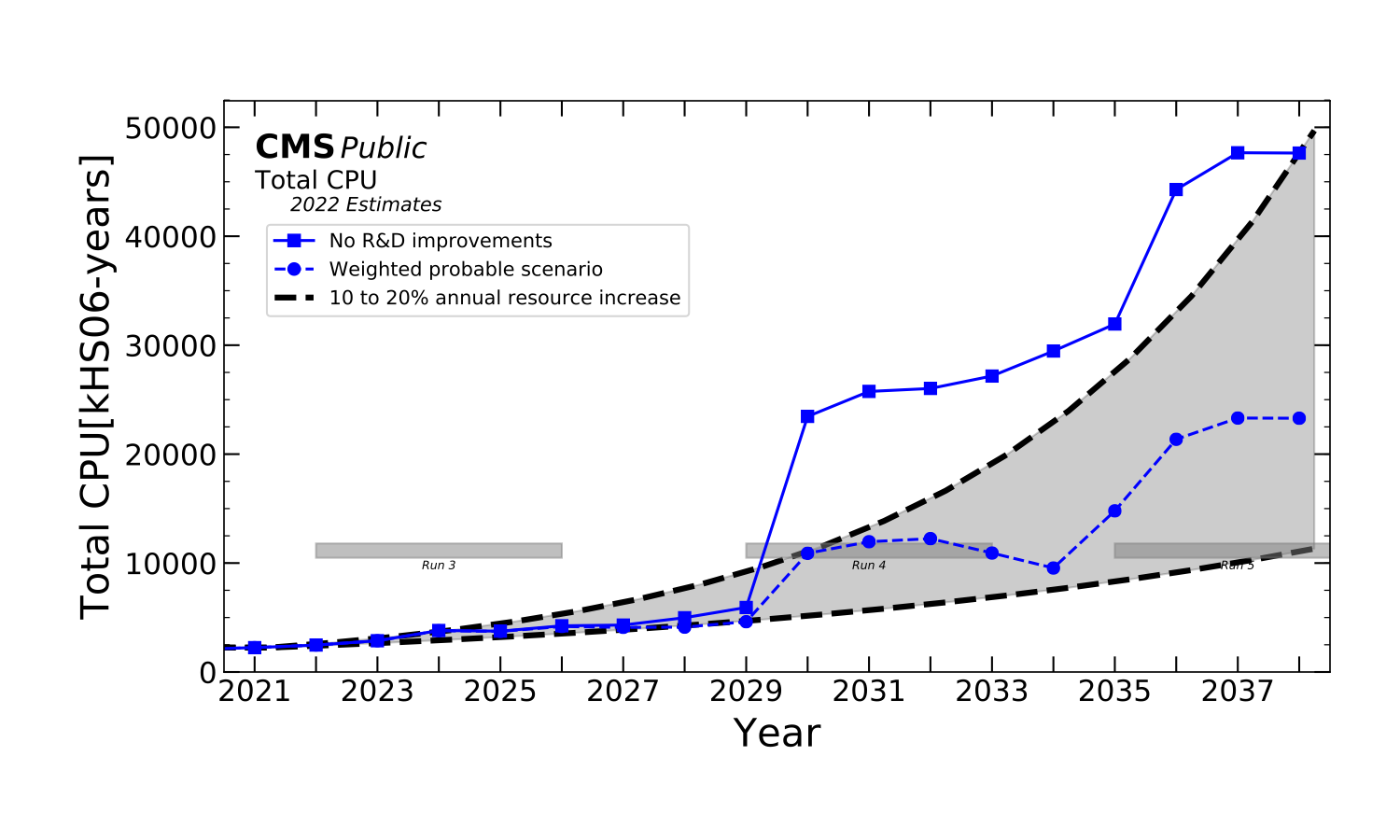}
\includegraphics[width=0.48\textwidth]{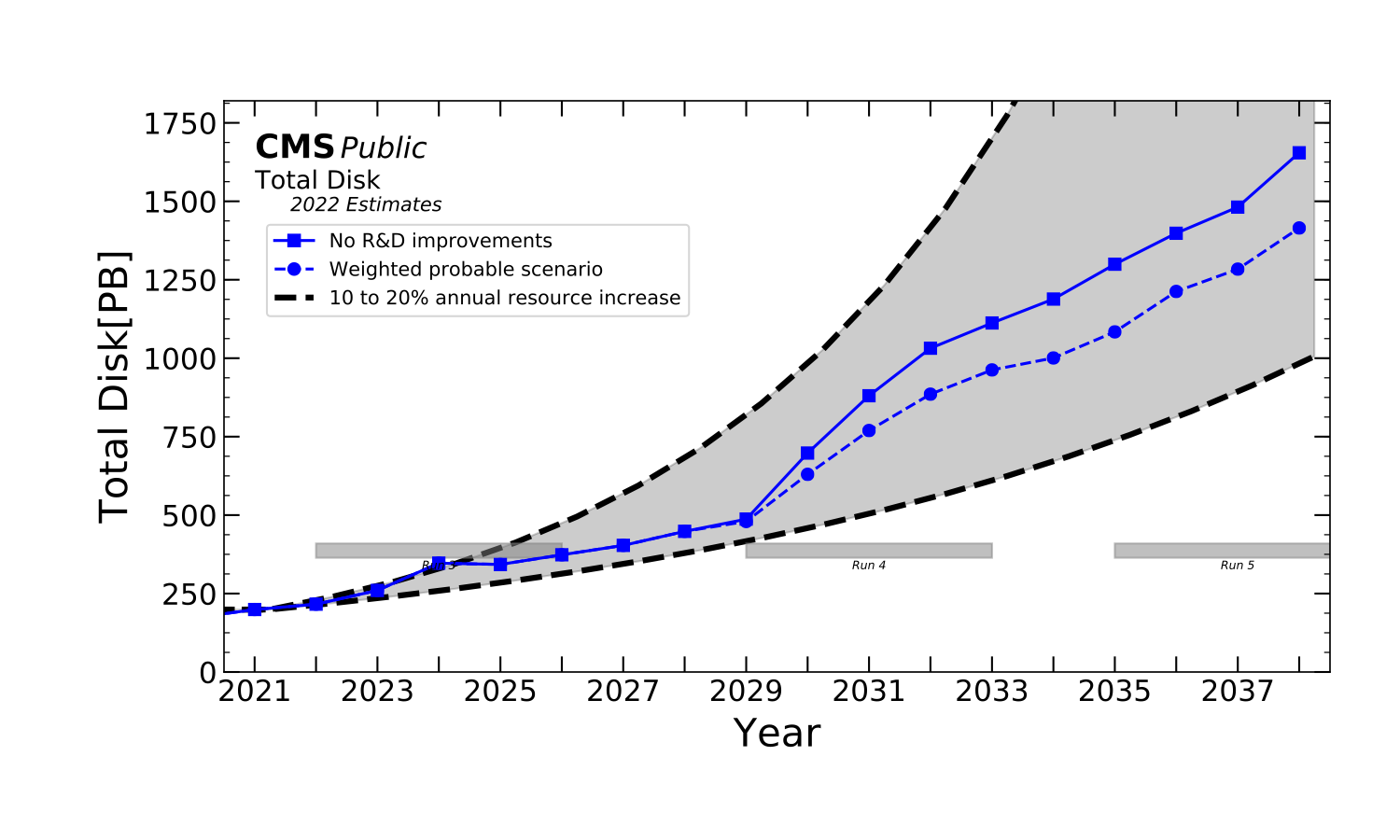} \\
\includegraphics[width=0.48\textwidth]{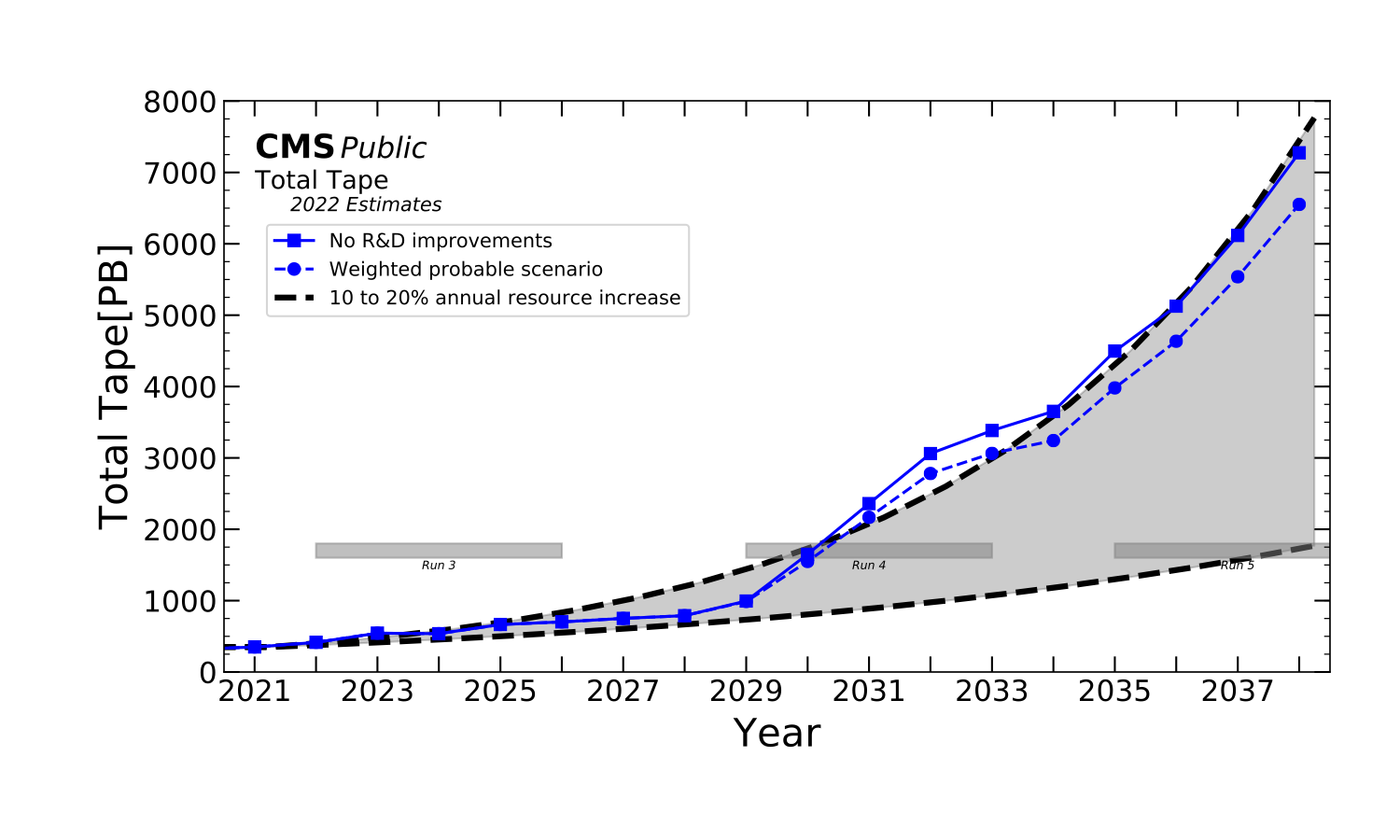}
  \caption{Estimated computing resource needs for CMS~\cite{bib:cmsofflinecomputingresults}. Shown are the modeled annual projections of total CPU and disk needs for CMS through Run 4. The estimated needs for each computing model scenario are shown by the blue lines. The gray band shows the projected resource availability for an example scenario that extrapolates the 2021 CMS pledged resources using an annual increase in available resources of between 10\% and 20\%. This assumes current WLCG cost projections~\cite{campana_simone_2021_5499655} and a warranty + 3 years replacement cycle of hardware.}
\end{figure}

These baseline scenarios still carry a variety of uncertainties and risks. Improvements in equipment pricing may not be as favorable as in the model~\cite{campana_simone_2021_5499655}. It may not be possible to reduce the number of promptly reconstructed events as much as planned. Fast simulation may not be sufficient for as much of the physics analyses as we predict, leading to a greater need for processing resources. Not enough analyses may be able to switch to the most compact data format for analysis (NanoAOD~\cite{Ehataht:2020ebp,Petrucciani_2015}), leading to greater resource demands for processing and storage.

To protect against these risks, the U.S.~CMS Operations Program has started a focused program of research that aims to meet the computing challenges of the HL-LHC and reduce resource needs further. The goals for these R\&D efforts are
\begin{enumerate}[(1)]
  \item to reach the required level of scalability for computing facilities, 
  \item to modernize software to efficiently use modern computing architectures and accelerators and improve software maintainability, 
  \item to identify opportunities for more disruptive changes in software and data analysis systems to enhance physics outcomes, 
  \item to minimize technology risks, and 
  \item to improve on needs for maintenance and operations efforts.
\end{enumerate}
The LHC Run~3 that started in 2022  provides opportunities for a "dress rehearsal" of many new computing techniques.

\section{Grand Challenges for HL-LHC Computing and Software}
\label{sec:challenges}

In order to effectively focus and structure the U.S.~CMS R\&D efforts, we organize the innovation, research, and development needs for HL-LHC computing into the four \emph{Grand Challenges} that encompass the advances needed for HL-LHC computing to succeed:

\begin{enumerate}[(1)]
\item \textbf{Modernizing Physics Software and Improving Algorithms}

Exploit novel algorithms, including ML/AI, reduce algorithmic complexity, increase computational intensity, and provide core software infrastructure to enable effective use of modern hardware and accelerators. The work is organized in the following work packages:

\begin{itemize}
    \item Core Software Framework and Software Portability
    \item Establish Performance Metric and Performance Baseline for Physics Software
    \item U.S. Contributions to the Charged Particle Tracking Software
    \item U.S. Contributions to Software for High Granularity Calorimeter
    \item U.S. Contributions to CMS Advanced Algorithms Work
\end{itemize}

\item \textbf{Building Infrastructure for Exabyte-Scale Datasets}

Build infrastructure to archive, store, transfer, and provide access to exabyte-scale datasets. Explore data lakes and custodial storage: establish a technology and cost model for custodial/archival storage facilities which manages operations costs, and optimizes hardware costs. Orchestrate computational services and data access, provide intelligent network services, and master the integration of national and international computing resources, including High Performance Computing centers (HPCs). Work packages are:

\begin{itemize}
    \item Produce Exabyte-scale Central Datasets
    \item Manage Exabyte-scale Network Flows
    \item Store Exabyte-scale Central Datasets
    \item Derive Petabyte-size Analysis Datasets
    \item Provide access to Petabyte-size Analysis Datasets
    \item Operate Facilities to support Production, Storage and Transfer of Exabyte-scale Datasets
\end{itemize}

\item \textbf{Transforming the Scientific Data Analysis Process}

Re-imagine how physicists interact with data when extracting science: seize on industry advances in data science, consider required performance parameters like throughput, latency, ease of use, and functionality, and develop the facility and software infrastructure to support thousands of physicists analyzing exabytes of data. Work packages are:

\begin{itemize}
    \item Provide Column-Wise Analysis Software Infrastructure
    \item Provide Column-Wise Analysis Facility Infrastructure
    \item Provide Column-Wise Data Augmentation Infrastructure
\end{itemize}

\item \textbf{Transition from R\&D to Operations}

The R\&D program will contribute to several advances in infrastructure particularly in analysis facilities and networking/storage that will need to be rolled into operations. This transition needs to include adequate testing, monitoring, and training/documentation for a successful adoption by the operational staff. Work packages are:

\begin{itemize}
    \item Develop sufficient Monitoring Capabilities
    \item Assess Readiness of Systems/Services for HL-LHC for Transition to User Testing and Operation
    \item Write Documentation for Systems/Services
    \item Provide Training for Operators and Users
\end{itemize} 

\end{enumerate}

\section{U.S. and International Partners}
\label{sec:partners}

The U.S.~CMS Operations Program partners in these endeavors with its host lab, Fermilab, its partner U.S. institutes and a number of organizations and consortia, allowing leverage of expertise and to find synergies between projects. 

U.S.~CMS has been proactive in developing an ecosystem for computing and software related research and developments, working with U.S. and international partners. The ecosystem includes {\em research partnerships}, like those with national labs and with CERN~\cite{CERN}; national and international {\em consortia} like the Open Science Grid (OSG)~\cite{osg} and the HEP Software Foundation (HSF)~\cite{hsf}; joint and {\em collaborative projects}, many of which are listed below, and {\em community efforts} like the joint Blueprint activities with U.S.~ATLAS, OSG, DOE Energy Science Network (ESnet)~\cite{esnet}, the Institute for Research and Innovation in Software for High Energy Physics (IRIS-HEP)~\cite{irishep}, and the Snowmass Computational Frontier~\cite{snowmasscomp}. This vast and proactively developed ecosystem and our partnerships provide strength, depth, and cohesiveness in the efforts to develop the HL-LHC computing environment for CMS.


We summarize below partnerships and collaborative work areas with external projects.

Within the DOE program the Center for Computational Excellence (HEP-CCE)~\cite{hepcce} is an important partner, that brings together the expertise of HEP physicists and applied math and computational scientists from DOE's Advanced Scientific Computing Research (ASCR)~\cite{ascr} program in an alliance of four national labs (Fermilab, BNL~\cite{bnl}, ANL~\cite{anl}, LBNL~\cite{lbnl}). The three-year project focuses on the current and future exascale HPC systems and enabling HEP experiments to efficiently exploit their capabilities. It supports research in areas of interest that are shared by multiple experiments and leverages the respective strengths of the laboratories. The highest priority for U.S.~CMS is to explore techniques that enable HEP software algorithms to be efficiently executed on a variety of advanced computing hardware architectures. These portability techniques will allow the same code to be executed on CPUs, GPUs and other future accelerators, reducing the effort to evolve the code to new and future HPC machines. Another priority is the exploration of advanced I/O capabilities of the HPC systems and how to optimally use them from the experimental frameworks for HEP workflows.

DOE also funded two relevant projects under SciDAC-4 -- \emph{HEP Data Analytics on HPC}~\cite{scidac4-analytics} and \emph{HEP Event Reconstruction with Cutting Edge Computing Architectures}~\cite{bib:scidac4-reco}, both of which are led by Fermilab. The former is developing tools and techniques to take advantage of HPC resources in analysis. The latter is developing vectorized and GPU-based tracking techniques for use in both the trigger and reconstruction.

The LHC program relies on very high throughput transatlantic networks as part of its international computational and data infrastructure. ESnet is both provider for the networking infrastructure with CERN and other international computing centers, and a partner for the domestic networking and high-throughput virtual circuits with partner universities (LHCONE~\cite{lhcone}). Increasingly, the LHC experiments make intense use of HPC centers, and we engage in setting requirements, and plan  the network access and integration of HPC centers into the LHC data infrastructure with ESnet. A focus for HL-LHC will be adding networks as another resource that is managed. Apart from processing and storage resources, the goal is to add the management of networking resources to the portfolio of the experiments to optimize science throughput.

At the same level, the NSF-funded IRIS-HEP contributes partnerships in vital R\&D activities. IRIS-HEP serves as an active center for software research, functions as an intellectual hub for the larger HEP community-wide software research efforts, and will transform the operational services required to ensure the success of the HL-LHC scientific program. In particular, U.S.~CMS is working with IRIS-HEP on the development of innovative algorithms for data reconstruction and triggering; development of data organization, management, and access systems for the community’s upcoming Exascale era; and development of highly performant analysis systems that reduce "time-to-insight" and maximize the HL-LHC physics potential.

In addition, IRIS-HEP funds the parts of the OSG that provide critical infrastructure services for the LHC experiments. OSG has partnered with U.S. LHC over many years in deploying novel computing solutions across the distributed high-throughput computing system and ensuring that they can be operated at sufficient scale. This partnership will continue through the HL-LHC era. The HEP Software Foundation (HSF) and the Worldwide LHC Computing Grid (WLCG)~\cite{wlcg} have also been important international partners in research and will continue to be so. 

\section{Summary}
\label{sec:execsum}

U.S.~CMS, in partnership with its host lab Fermilab, its seven Tier-2 institutes and the U.S.~CMS research and national and international S\&C communities, has established a HL-LHC R\&D program to plan for and build the computing services required for full scientific exploitation of the CMS HL-LHC upgrades. The program directly addresses the four Grand Challenges of HL-LHC software and computing: (1) Modernizing Physics Software and Improving Algorithms, (2) Building Infrastructure for Exabyte-Scale Datasets, (3) Transforming the Scientific Data Analysis Process, and (4) Transition from R\&D to Operations. It is fully integrated in the CMS plans for HL-LHC and delivers important components to the overall system.

Management and oversight is provided as part of the on-going U.S.~CMS operations program. It also contributes to and strongly leverages  the wider computing efforts at Fermilab, the U.S. universities participating in U.S.~CMS S\&C activities, and the Open Science Grid. It is participating in HEP-CCE within the DOE lab complex, and in the NSF-funded IRIS-HEP LHC Software Institute. In addition, it works closely with the HEP Software Foundation and the CERN-based WLCG program.
\section{Acknowledgments}
This manuscript has been authored by Fermi Research Alliance, LLC under Contract No. DE-AC02-07CH11359 with the U.S. Department of Energy, Office of Science, Office of High Energy Physics, and by the National Science Foundation under Cooperative Agreement PHY-2121686.

%
%

\end{document}